\documentclass[10 pt,final,journal,letterpaper,oneside,doublecolumn]{IEEEtran}
\usepackage{multicol}
\usepackage{cite}
\usepackage{multirow}
\usepackage[pdftex]{graphicx}
\usepackage[lined,boxruled]{algorithm2e}
\usepackage{algorithmic}
\usepackage[small]{caption}
\usepackage{float}
\usepackage[utf8]{inputenc}
\usepackage{amsmath, amsthm, amssymb}

\usepackage[english]{babel}
\usepackage{amsthm}
\usepackage{xspace}
\usepackage{fancyhdr}
\usepackage{amssymb,amsmath}
\usepackage{subfig}
\usepackage{textcomp}
\usepackage{epstopdf}
\usepackage{bbding}
\usepackage{color}
\usepackage{pifont}

\begin{document}
%
\title{\huge NOMA-enabled Backscatter Communications for Green Transportation in Automotive-Industry 5.0}

\author{Wali Ullah Khan, Asim Ihsan, Tu N. Nguyen, Zain Ali, and Muhammad Awais Javed\thanks{W. U. Khan is with Interdisciplinary Centre for Security, Reliability and Trust (SnT), 1855 Luxembourg City, University of Luxembourg (email: waliullahkhan30@gmail.com).

A. Ihsan is with the Department of Information and Communication Engineering, Shanghai Jiao Tong University, Shanghai 200240, China (email: ihsanasim@sjtu.edu.cn).

Tu N. Nguyen is with the Department of Computer Science, Purdue University Fort Wayne, Fort Wayne, IN 46805, USA (email: nguyent@pfw.edu).

Z. Ali is with the Department of Electrical and Computer Engineering, University of California, Santa Cruz, USA (email: zainalihanan1@gmail.com).

M. A. Javed is with the Department of Electrical and
Computer Engineering, COMSATS University Islamabad, Islamabad 45550, Pakistan (email: awais.javed@comsats.edu.pk).

}}%

\maketitle

\begin{abstract}
Automotive-Industry 5.0 will use emerging 6G communications to provide robust, computationally intelligent, and energy-efficient data sharing among various onboard sensors, vehicles, and other Intelligent Transportation System (ITS) entities. Non-Orthogonal Multiple Access (NOMA) and backscatter communications are two key techniques of 6G communications for enhanced spectrum and energy efficiency. In this paper, we provide an introduction to green transportation and also discuss the advantages of using backscatter communications and NOMA in Automotive Industry 5.0. We also briefly review the recent work in the area of NOMA empowered backscatter communications. We discuss different use cases of backscatter communications in NOMA-enabled 6G vehicular networks. We also propose a multi-cell optimization framework to maximize the energy efficiency of the backscatter-enabled NOMA vehicular network. In particular, we jointly optimize the transmit power of the roadside unit and the reflection coefficient of the backscatter device in each cell, where several practical constraints are also taken into account. The problem of energy efficiency is formulated as nonconvex which is hard to solve directly. Thus, first, we adopt the Dinkelbach method to transform the objective function into a subtractive one, then we decouple the problem into two subproblems. Second, we employ dual theory and KKT conditions to obtain efficient solutions. Finally, we highlight some open issues and future research opportunities related to NOMA-enabled backscatter communications in 6G vehicular networks.
\end{abstract}


\IEEEpeerreviewmaketitle

\section{Introduction}
Automotive-Industry 5.0 will focus on the smart interaction between humans and autonomous vehicles. Industry 4.0 was about smart automation using the machine to machine connectivity, intelligent computing, and big data analytics. Internet of Things (IoT) and machine learning were the major enablers of Industry 4.0 \cite{vehicular_1}. In the context of the automotive industry, Industry 4.0 improved the quality and scalability of vehicle production and provided applications such as passenger safety, smart traffic management, and infotainment. Industry 4.0 enabled vehicles were equipped with various sensors to detect nearby vehicles, share data with the neighborhood, and connect with the cloud. Intelligent big data analytics applied on traffic and vehicle health data ensured vehicles were serviced within time, and choose fuel-efficient and undamaged travel routes \cite{vehicular_2}.

Industry 5.0 will be empowered by upcoming technologies such as 6G communications \cite{javed_6g}, artificial intelligence techniques \cite{ai_1}, fog computing \cite{fog_1, fog_2}, and blockchain \cite{bchain_1}. In the automotive sector, Industry 5.0 will enhance the robustness of wireless connectivity, allow humans to share data in collaboration with vehicles, and develop reliable autonomous driving applications \cite{9345447}. Moreover, Automotive-Industry 5.0 will focus on preparing an ecosystem for green transportation with minimum carbon dioxide emissions. 6G communications will be a key enabler of Automotive-Industry 5.0 as it will provide robust and high-speed communications required to meet the needs of large data sharing between vehicle sensors and the cloud \cite{5g_1, 5g_2}. Besides, 6G will also support energy-efficient green communications \cite{energy_1}.

With the increase in the number of vehicles on the road, the consumption of fossil fuels, carbon emission and noise pollution increases significantly. Moreover, 6G enabled electric vehicles will be a major component of Automotive Industry 5.0 and will share huge mobility and traffic related data with each other. This data will be generated and shared by many small and power constrained sensors placed inside the vehicles, and battery operated infrastructure units known as Road Side Units (RSUs). Backscatter communication and Non-Orthogonal Multiple Access (NOMA) are two promising 6G technologies that can improve the energy efficiency of IoV, thus realizing the green transportation goal \cite{wu2019energy,8368232}.

Backscatter communication uses the existing RF resources to transfer data between vehicles and RSUs in the Internet of Vehicles (IoV) network \cite{8368232}. Thus, backscatter communication can extend the lifetime of vehicle sensors and RSUs by reflecting the existing RF signals towards intended vehicles without exploiting any oscillatory circuity. 

NOMA is a promising air interface technique in which power-domain is used for multiple access as compared to previous air interface techniques that relied on other domains, i.e., time, frequency, and code \cite{8861078}. In NOMA, multiple vehicles can access the same spectrum/time resource for communication \cite{zhang2021centralized}. More specifically, multiple vehicles can be multiplexed through different power levels by using the superposition coding technique at the transmitter side. Then, every vehicle can decode its desired signal by using the successive interference cancellation (SIC) technique.

This paper first provides an overview of green transportation in Automotive Industry 5.0 and significance of two key 6G technologies i.e., backscatter communications and NOMA for green transportation applications. Then we discuss the recent developments in NOMA-enabled backscatter communication. We also present different use cases of NOMA-aided backscatter for 6G V2X communications. Moreover, we propose a new optimization framework of NOMA-enabled backscatter communication for multi-cell IoV networks. Specifically, we jointly optimize the transmit power of RSU and reflection coefficient of backscatter tag in each cell to maximize the total achievable energy efficiency of NOMA enabled backscatter based IoV network. Finally, open issues and some exciting future research opportunities for designing backscatter based massive green V2X systems using NOMA are indicated, followed by concluding remarks.


\begin{table*}
\centering
\caption{A summary and comparison of backscatter communication frameworks involving NOMA with our proposed solution}
\begin{tabular}{|p{0.5cm}|p{1.2cm}|p{2.6cm}|p{1cm}|p{1.1cm}|p{9cm}|} \hline
Ref.& Scenario & Work Objective & Type & SIC & Proposed Solution \\ \hline\hline
\cite{9345447}& single-cell & spectral efficiency & Ambient & Perfect & Proposed a joint optimization framework of power allocation at BS and RSU to maximize the sum rate of backscatter-enabled NOMA V2X communication. \\ \hline
\cite{9131891}& single-cell & Physical layer security & Ambient & Perfect & Derived closed-form expressions of outage and intercept probabilities for the physical layer security of NOMA-enabled backscatter communication.\\ \hline
\cite{8636518} & single-cell & Ergodic capacity & Symbiotic & Perfect & Derived closed-form expressions for ergodic capacity and outage probability of NOMA-enabled backscatter communication. \\ \hline \cite{9319204}& single-cell & Physical layer security & Ambient & Imperfect & Derived closed-form expressions for outage and intercept probabilities of NOMA-enabled backscatter communication under imperfect SIC/CSI.\\ \hline
\cite{le2019outage}& single-cell & Outage performance & Ambient & Perfect & Derived closed-form expression for the outage probability of NOMA-enabled MISO backscatter communication. \\ \hline
\cite{9328505} & single-cell & Spectral efficiency & Ambient & Imperfect & Provided joint optimization framework of power and reflection coefficient to maximize the sum capacity of NOMA-enabled backscatter communication.\\ \hline
\cite{9223730}& single-cell & Energy efficiency & Ambient & Perfect & Optimized transmit power and reflection coefficient to maximize the energy efficiency of NOMA-enabled backscatter communication.\\ \hline
\cite{9162938}& single-cell & Physical layer security & Ambient & Perfect & Presented an optimization framework to maximize the sum secrecy rate of NOMA-enabled backscatter communication under multiple eavesdroppers.\\ \hline
\cite{8851217}& single-cell & System throughput & Bistatic & Perfect & Provided a joint optimization of time and reflection coefficient to maximize the minimum throughput of NOMA-enabled backscatter communication.\\ \hline
\cite{9122620}& single-cell & Ergodic capacity & Ambient & Perfect & Derived the optimal reflection coefficient, total power budget of BS, and power allocation coefficient of users to maximize the ergodic capacity of backscatter communication. \\ \hline
\cite{8962090}& single-cell & System throughput & Symbiotic & Perfect & Investigated a joint optimization of time, transmit power and reflection coefficient to maximize the minimum throughput of backscatter communication. \\ \hline
This work & Multi-cell & Energy efficiency & Ambient & Perfect  \& imperfect & We provide a multi-cell optimization framework that jointly optimizes the transmit power of RSU and reflection coefficient of backscatter device in each cell. The objective is to maximize the total achievable energy efficiency of NOMA vehicular network under perfect and imperfect SIC detecting. \\ \hline
\end{tabular}
\label{tab2}
\end{table*}

\section{NOMA enabled Backscatter Communications in Industry 5.0}
In this section, we first provide an overview of green transportation in Automotive Industry 5.0 and significance of NOMA enabled backscatter communications for next generation automotive applications. Then, we present some recent developments in NOMA-enabled backscatter communication systems.


\subsection{Green Transportation in Automotive Industry 5.0}
Green transportation refers to techniques and technologies that enhance the fuel efficiency, safety and sustainability of transportation system. The major aim of green transportation is to reduce the reliance on petroleum fuelled vehicles that result in large carbon emissions and pollution. By finding alternate environmental friendly energy resources to fuel different means of transport, green house gas emissions can be significantly reduced. 

Automotive Industry 5.0 will lay emphasis on green transportation to improve the environment and move towards a sustainable and autonomous system. Smart techniques such as power-aware wireless communication algorithms (such as NOMA), intelligent resource allocation procedures and cooperative computing will be used to minimize the energy consumption of IoV network. Moreover, innovative technologies such as energy efficient sensors, backscatter communications, intelligent reflecting surfaces will be used to achieve low power system operation.

\subsection{Significance of backscatter communications and NOMA in Automotive Industry 5.0}

6G communications will play a major role in realizing green transportation system in Automotive Industry 5.0. A major use case of 6G in future automotive industry will be efficient traffic management application. Traffic jam and congestion has been observed to be a major source of carbon emissions \cite{cong_1}. With 6G communications, wireless data sharing among ITS entities such as vehicles, infrastructure Road Side Units (RSUs) and city wide traffic data centers can be enabled. As a result, large accurate mobility data set can be collected and efficient and congestion free traffic route management can be implemented by applying machine learning algorithms.

Backscatter communication will facilitate in energy efficient communications for traffic management applications by providing ultra-low power transmission. Utilizing the existing RF signal, backscatter communication can harvest energy for circuit operation, data modulation, and signal reflection. In general, a backscatter communication system consists of a passive backscatter device and a reader. Backscatter communication can be categorized into three different types, i.e., monostatic backscatter communication, biostatic backscatter communication, and ambient backscatter communication \cite{9261963}. As RSUs will be battery operated, their energy efficiency will be critical. By installation of backscatter tags on the vehicles that act as reflectors, energy consumed by RSUs can be significantly reduced.

NOMA is an efficient multiple access technique that provides adaptive transmit power allocation to the vehicles based on their channel conditions. According to the NOMA protocol, a vehicle with comparatively good channel conditions is assigned lower transmit power as compared to a vehicle with bad channel conditions \cite{patel2021performance}. This also guarantees the quality of services (QoS) for vehicles with weak channel conditions. It is important to note that the vehicle with less transmit power needs to apply SIC to subtract the high power signal of the weak vehicle from the received signal. However, the weak vehicle cannot apply SIC and decode the signal by treating the signal of the strong vehicle as interference \cite{9261140}.

\subsection{Recent Development in NOMA enabled Backscatter Communication}
Recently, NOMA has been integrated into backscatter communication. For instance, the authors of \cite{9131891} investigated the reliability and security of NOMA-enabled backscatter communication with in-phase and quadrature-phase imbalance. Their objective was to derive the closed-form expressions for the outage and intercept probabilities. Zhang {\it et al.} \cite{8636518} derived the closed-form expressions for the ergodic capacity and outage probability in a two-user NOMA-enabled symbiotic system. To improve the physical layer security, the work in \cite{9319204} derived the outage and intercept probabilities of backscatter communication under residual hardware impairments, channel estimation error, and imperfect SIC decoding. They also proposed artificial noise which acted as a jammer to the malicious eavesdropper. Le {\it et al.} \cite{le2019outage} investigated the outage performance of multiple-input single-output NOMA-enabled backscatter communication. They derived a closed-form expression for the outage probability of the system.

Khan {\it et al.} \cite{9328505} proposed a joint optimization framework for NOMA-enabled backscatter communication under imperfect SIC decoding. Their objective was to maximize the sum capacity by optimizing the transmit power of the source and the reflection power of the backscatter tag. Research of \cite{9223730} proposed a joint energy-efficient optimization framework for NOMA-enabled backscatter communication with quality of services guarantee. Reference \cite{9162938} provided a secrecy rate maximization problem in NOMA-enabled backscatter communication under multiple non-colluding eavesdroppers. Yang {\it et al.} \cite{8851217} investigated a joint optimization problem of time and power reflection coefficient to maximize the minimum throughput of NOMA-enabled multiple backscatter communication networks. Similarly, in another study \cite{8962090}, Liao {\it et al.} provided a joint optimization framework of transmit power at the source, time, and power reflection coefficients to maximize the minimum throughput of NOMA-enabled multiple backscatter communication networks. Chen {\it et al.} \cite{9122620} derived optimal power budget at BS, power allocation coefficient of users, and reflection coefficient of backscatter tag to maximize the ergodic capacity of backscatter communication. Of late, the authors of \cite{9345447} investigated the problem of spectral efficiency for backscatter-enabled NOMA V2X communication. A complete summary of the above discussed NOMA-enabled backscatter communication framework is shown in Table I.

\section{Use Cases of NOMA-enabled Backscatter Communication in 6G IoV}
In 6G-based IoV, the proliferation of vehicles and other connected entities limit the availability of energy for connecting them through different wireless technologies. IoVs contain vehicles that need to maintain reliable communications with each other at different tiers such as Vehicle-to-RSU (V2R), Vehicle-to-Backscatter Sensor (V2BS) Vehicle-to-Infrastructure (V2I), Vehicle-to-Vehicle (V2V), Vehicle-to-Pedestrian (V2P), Vehicle-to-Bicycle (V2B), Vehicle-to-UAV (V2U) and Backscatter Sensor-to-Everything (BS2X) as shown in Figure 1. Passive backscatter communication technology provides an energy-efficient and practical alternative to use radio devices on existing sensor technologies. By combining power domain NOMA with backscatter communications, several green transportation use cases and applications for 6G IoV can be realized. We discuss three possible use cases in the following subsections.   


\subsection{V2I Case}
In cellular vehicle-to-everything (C-V2X), infrastructure-based communication (V2I/I2V) is carried out through the cellular interface. It plays the role of a coordinator by collecting real-time local and global information for smart transport management, logistics, navigation, and environmental monitoring. It also provides real-time safety messages to the vehicles, based on condition and location, such as traffic jam warning, safe distance warning, speed limit information, lane-keeping, intersection safety, and accident warning. V2I/I2V delivers various services to all users in IoV through both eNodeB and RSU. RSUs are deployed on the roads to manage heavy data traffics, which reduces severe data congestion in cellular networks.

NOMA can be used for the exchange of information between infrastructure and other users in V2X, which may achieve more efficient energy and spectrum resource utilization \cite{9345447}. In the future IoV networks, there will be massively connected sensors; therefore it demands passive sensors that can be wirelessly powered through RF energy harvesting or any other technology such as piezoelectric materials technology \cite{9000613}. Backscatter-aided roadside sensors can be wirelessly operated by using the signals of Base Station (BS) to reflect their sensed information to the vehicles in a downlink scenario \cite{9345447}. Besides, dedicated carrier emitters can be installed on the roadsides to wirelessly power the roadside backscatter sensors for sending the sensed information to the nearest RSU in the uplink scenario.
\begin{figure*}[!t]
\centering
\includegraphics [width=0.75\textwidth]{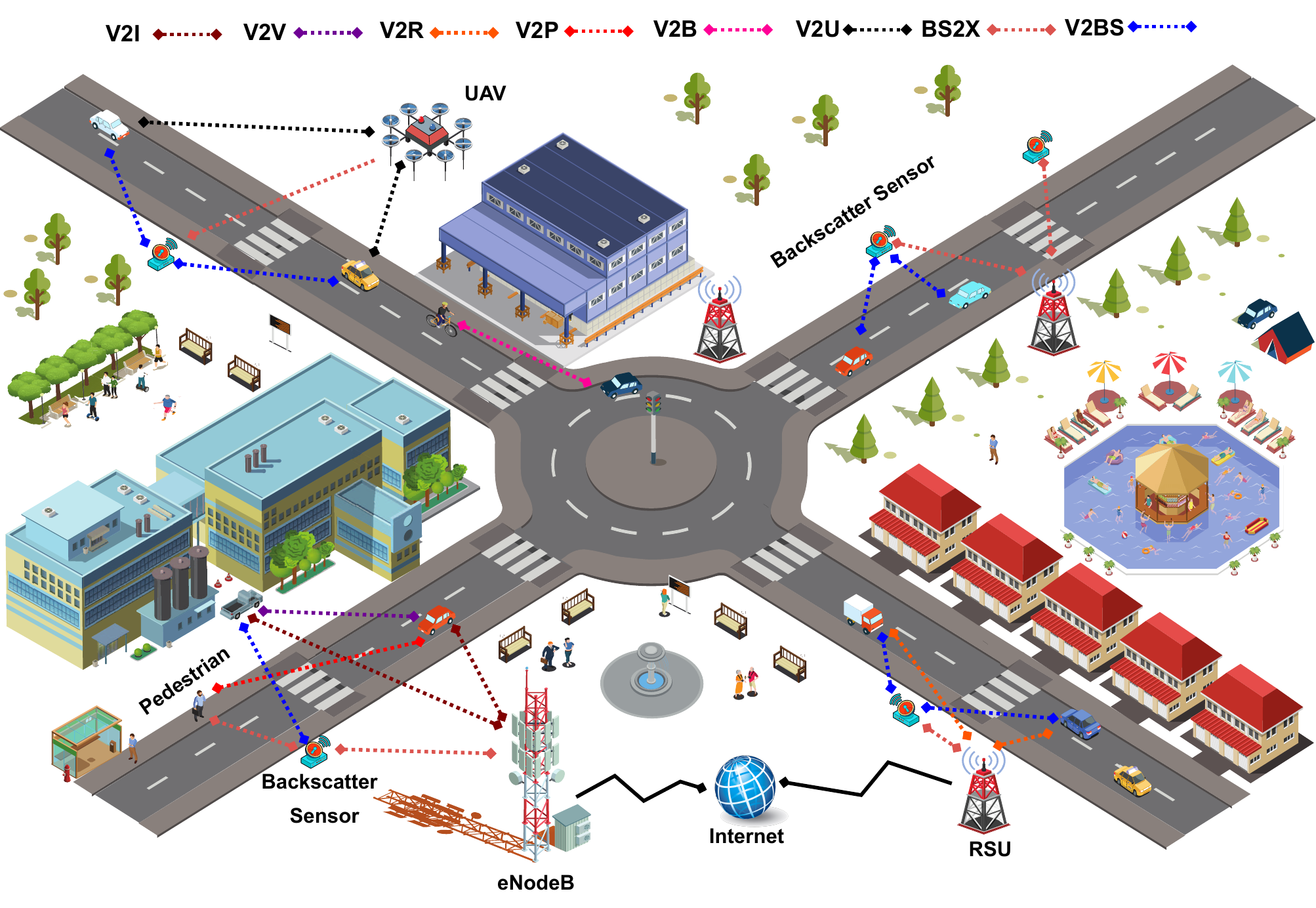}
\caption{Use case of backscatter-aided NOMA in 6G V2X communications which include Vehicle-to-RSU (V2R), Vehicle-to-Backscatter Sensor (V2BS) Vehicleto-Infrastructure (V2I), Vehicle-to-Vehicle (V2V), Vehicleto-Pedestrian (V2P), Vehicle-to-Bicycle (V2B), Vehicle-to-UAV (V2U) and Backscatter Sensor-to-Everything (BS2X).}
\label{Fig2}
\end{figure*}

\subsection{V2V Case}
In C-V2X, V2V communication is necessary for safety applications and is performed through the PC5 interface. The vision behind V2V research is to make each vehicle on the road, including motorcycles and bicycles, capable to communicate with each other. Then, the exchange of a rich set of data between them through the NOMA principle will efficiently support the next generation of safety applications and systems. Different from the orthogonal multiple access technologies, in NOMA-enabled V2V communication, a lead vehicle which act as a transmitter can broadcast the superimposed messages to multiple vehicles over the same spectrum at the same time \cite{8755879}. This feature significantly improves the performance of V2V communication. Smart sensing technologies are the demand of future IoV networks. Today’s latest vehicles are already equipped with sensors such as lidars, radars, ultrasonic sensors, and cameras that detect objects around the vehicles and provide safety features such as forward collision detection, blind-spot detection, and lane change assistance. These large numbers of sensors installed on vehicles and roads can be wirelessly powered through ambient sensors along with backscatter communication for energy-efficient V2V communications \cite{8809655}.

\subsection{UAV Case}
Integration of NOMA, backscatter, and UAV-assisted wireless communication into the V2X network has great potential to provide energy and spectral efficient communication to the moving vehicles with larger coverage. UAV can assist V2X communications either as a flying base station or as a relaying node. The performance of UAV-assisted V2X communication can be further improved by implementing NOMA, which provides improved sum-rate and connectivity to UAVs and vehicles \cite{uav_1}. Besides, Backscatter communication can aid UAV-enabled V2X communications in tackling lifetime energy management problems of massive sensors used for various IoV applications. UAV can act as carrier emitter as well as data transmitter and collector for backscatter communications. Thus UAVs can support energy efficient coverage range extension of backscatter tags by collecting sensed information from these tags and quick data delivery to the base station by using high mobility of UAVs \cite{9222571, uav_2}. 
\begin{figure*}[!t]
\centering
\includegraphics [width=0.70\textwidth]{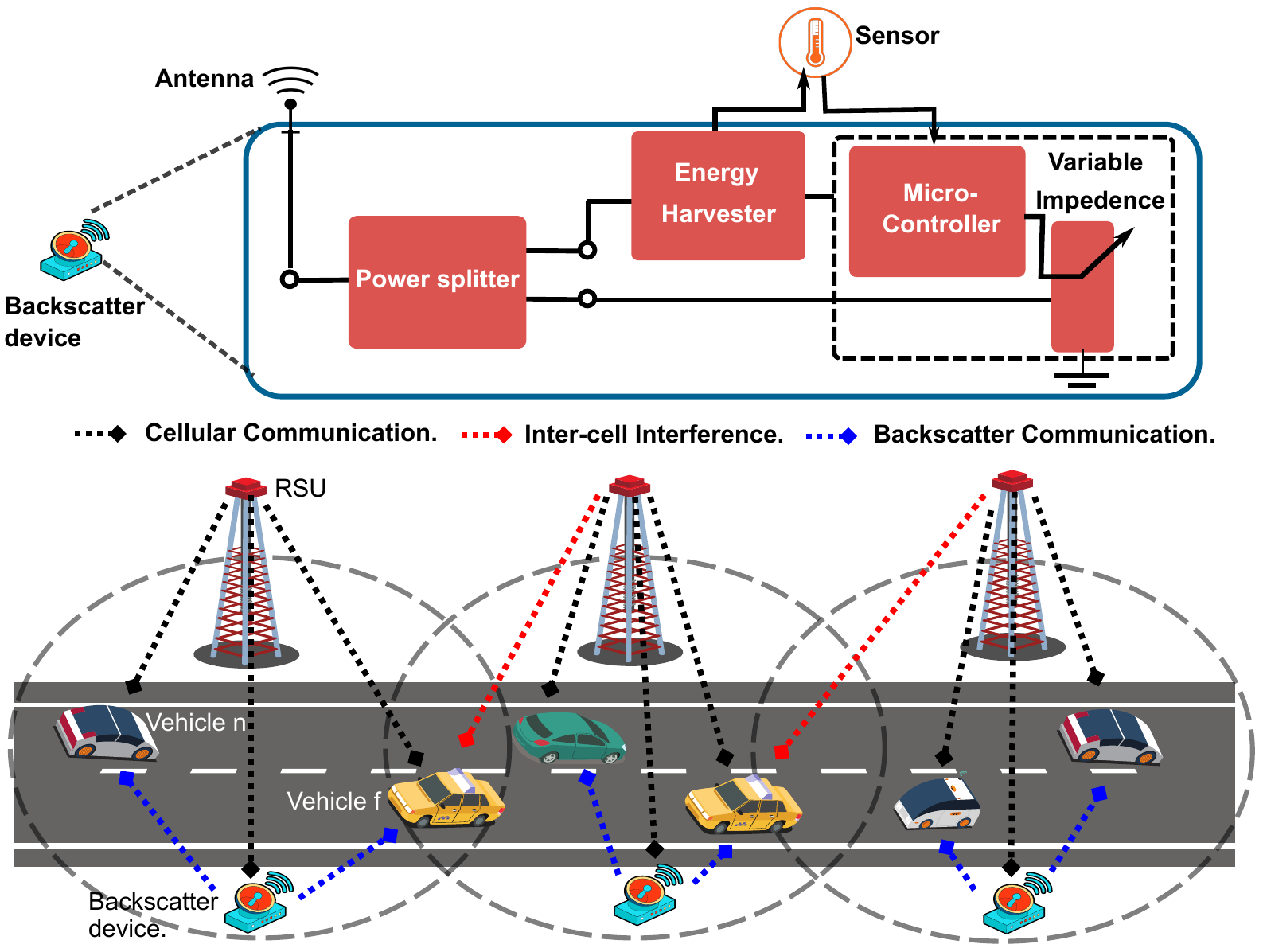}
\caption{ NOMA-enabled backscatter communication for multi-cell vehicular networks. In each cell, an RSU serves two vehicles using downlink NOMA protocols. All the RSUs share the same spectrum resources and receive cochannel interference from each other. Meanwhile, the backscatter device also receives the superimposed signal of RSU, uses it to modulate its message and reflect it towards vehicles. Our aim aim is to maximize the total achievable energy efficiency of backscatter-enabled NOMA vehicular network under perfect and imperfect SIC detecting. }
\label{Fig3}
\end{figure*}
\section{Energy-efficient Optimization Framework for Backscatter-enabled NOMA Vehicular Networks under Imperfect SIC}
In this section, we first discuss the system and channel models followed by different steps of problem formulation along with various practical constraints. Then we provide a proposed energy-efficient solution based on Dinkelbach method and KKT conditions.  
\subsection{System Model and Problem Formulation}
As shown in Figure \ref{Fig3}, we consider a single-carrier multi-cell vehicular network, where an RSU in each cell serves two vehicles using downlink NOMA protocol\footnote{Following the recent literature on vehicular communications \cite{8809665,9345447}, the mobility of vehicles is not considered because it simultaneously changes the channel characteristic of vehicles which requires various vehicle association and admission control policies. Thus, we assume the perfect information of vehicles to focus on power allocation of RSUs and the reflection coefficient of backscatter devices in this study. The study of considering mobility and imperfect channel information is set aside for the future.}. Considering more than two vehicles over a single-carrier significantly increases the hardware complexity and processing delay of SIC detecting process. A vehicle with strong channel condition is denoted as vehicle $n$, and the one with weak channel condition is stated as vehicle $f$, respectively. To enhance the connection density of vehicles in the network, all RSUs share the same spectrum resources. We consider a backscatter device (denoted as $k$) in each cell also receives the superimposed signal of RSU (denoted as $s$) and use the it to modulate its own information. it to harvest some energy to operate its circuit. More specifically, the backscatter device $k$ first harvests energy from superimposed signal to operate the circuit and then reflects the information towards vehicle $n$ and vehicle $f$. In practical scenarios, the vehicles can not always decode the desired information from the superimposed signal. Thus, in this work we have considered imperfect SIC decoding at the receivers. It is assumed that various devices in the network are equipped with omnidirectional antennas and the channels between different devices undergo Rayleigh fading. As mentioned earlier that the channel gain of vehicle $n$ is stronger than the channel gain of vehicle $f$, i.e., $|g_{n,s}|>|g_{f,s}|$. Then, the transmit power of both vehicles should satisfy as $p_s\alpha_{n,s}\leq p_s\alpha_{f,s}$, where $p_s$ is the transmit power of RSU $s$, $\alpha_{n,s}$ denotes the power allocation coefficient of vehicle $f$ and $\alpha_{f,s}$ is power allocation coefficient of vehicle $f$. This power allocation at RSU will ensure the SIC decoding process at vehicle $n$.
The signal to interference plus noise ratio of the vehicle $n$ and vehicle $f$ associated with RSU $s$ are given as:
\begin{align}
\gamma_{n,s}=\frac{p_s\alpha_{n,s}(|g_{n,s}|^2+\beta_{k,s}|g_{k,s}|^2|h_{n,k}|^2)}{p_s\alpha_{f,s}|g_{n,s}|^2\delta+I_{n,s}+\sigma^2},
\end{align} 
\begin{align}
\gamma_{f,s}=\frac{p_s\alpha_{f,s}(|g_{f,s}|^2+\beta_{k,s}|g_{k,s}|^2|h_{f,k}|^2)}{p_s\alpha_{n,s}(|g_{f,s}|^2+\beta_{k,s}|g_{k,s}|^2|h_{f,k}|^2)+I_{f,s}+\sigma^2},
\end{align} 
where $\beta_{k,s}$ denotes the reflection coefficient of the backscatter device $k$ which is located in the geographical area of RSU $s$. $|g_{k,s}|^2$ is the channel gains from the RSU $s$ to backscatter device $k$. Accordingly, $|h_{n,k}|^2$ and $|h_{f,k}|^2$ are the channel gains from backscatter device $k$ to vehicle $n$ and vehicle $f$. Further, $\delta$ is the imperfect SIC parameter such as $\delta=\mathbb E[|x-\hat x|]$, where $x-\hat x$ shows the difference between original signal and the detected signal. In addition, $I_{\iota,s}=\sum_{\substack{s'=1 \\ s'\neq s}}^S |g_{\iota,s'}|^2 p_{s'}$ is the interference from the neighboring RSUs to the vehicle $\iota$ being served by RSU $s$. $\sigma^2$ represents the variance of additive white Gaussian noise.

The objective function is defined as the summation of each RSU's energy efficiency which can be obtained as the ratio of sum-rate and total power consumption. The power consumption for each RSU includes the circuit power and transmission power on this RSU. The energy efficiency is maximized through joint optimization of transmit power of RSUs and reflection powers of backscatter devices. This can be achieved through solving an energy efficiency optimization problem which is subjected to the minimum data rate of vehicles, maximum transmit power of RSUs and reflection coefficient of backscatter devices. Mathematically, the energy efficiency problem can be formulated as:
\begin{figure*}[t]
$x=p^2_s(-((|g_{n,s}|^2+\beta_{k,s}|g_{k,s}|^2|h_{n,k}|^2)(|g_{f,s}|^2+\beta_{k,s}|g_{k,s}|^2|h_{f,k}|^2)(|g_{f,s}|^2+\beta_{k,s}|g_{k,s}|^2|h_{f,k}|^2))(1+\lambda_{n,s})(I_{n,s}+\sigma^2+|g_{n,s}|^2\delta p_s)+|g_{n,s}|^2+\beta_{k,s}|g_{k,s}|^2|h_{n,k}|^2(|g_{f,s}|^2+\beta_{k,s}|g_{k,s}|^2|h_{f,k}|^2)^2(1+\lambda_{f,s})(I_{n,s}+\sigma^2+|g_{n,s}|^2+\delta p_s)+(|g_{n,s}|^2+\beta_{k,s}|g_{k,s}|^2|h_{n,k}|^2)(|g_{f,s}|^2+\beta_{k,s}|g_{k,s}|^2|h_{f,k}|^2)(|g_{n,s}|^2+\sigma^2)(1+\lambda_{f,s})((I_{f,s}+\sigma^2)+(|g_{f,s}|^2+\beta_{k,s}|g_{k,s}|^2|h_{f,k}|^2)p_s)-(|g_{f,s}|^2+\beta_{k,s}|g_{k,s}|^2|h_{f,k}|^2)(|g_{n,s}|^2+\sigma^2)^2(1+\lambda_{f,s})(I_{f,s}+\sigma^2)+(|g_{f,s}|^2+\beta_{k,s}|g_{k,s}|^2|h_{f,k}|^2)p_s)).$\\
$y=p_s(I_{n,s}+\sigma^2+|g_{n,s}|^2\delta p_s)(-(|g_{n,s}|^2+\beta_{k,s}|g_{k,s}|^2|h_{n,k}|^2)(I_{f,s}+\sigma^2)(-2(|g_{f,s}|^2+\beta_{k,s}|g_{k,s}|^2|h_{f,k}|^2)(1+\lambda_{n,s}))+(|g_{f,s}|^2+\beta_{k,s}|g_{k,s}|^2|h_{f,k}|^2)(2+L_{n,s}+\lambda_{f,s}))+(|g_{n,s}|^2+\beta_{k,s}|g_{k,s}|^2|h_{n,k}|^2)(|g_{f,s}|^2+\beta_{k,s}|g_{k,s}|^2|h_{f,k}|^2)(|g_{f,s}|^2+\beta_{k,s}|g_{k,s}|^2|h_{f,k}|^2)(\lambda_{n,s}-\lambda_{f,k})p_s+2(|g_{f,s}|^2+\beta_{k,s}|g_{k,s}|^2|h_{f,k}|^2)(|g_{n,s}|^2\delta)(1+\lambda_{n,s})(I_{f,s}+\sigma^2+(|g_{f,s}|^2+\beta_{k,s}|g_{k,s}|^2|h_{f,k}|^2)p_s).$\\
$
z=((I_{n,s}+\sigma^2)+(|g_{n,s}|^2\delta p_s))((|g_{n,s}|^2+\beta_{k,s}|g_{k,s}|^2|h_{n,k}|^2)(I_{n,s}+\sigma^2)^2(1+\lambda_{n,s})+(|g_{f,s}|^2+\beta_{k,s}|g_{k,s}|^2|h_{f,k}|^2)(-(I_{n,s}+\sigma^2)(1+\lambda_{f,s})(I_{f,s}+\sigma^2+(|g_{f,s}|^2+\beta_{k,s}|g_{k,s}|^2|h_{f,k}|^2)p_s)+p_s((|g_{n,s}|^2+\beta_{k,s}|g_{k,s}|^2|h_{n,k}|^2)(I_{f,s}+\sigma^2)(1+\lambda_{n,s})-|g_{n,s}|^2\delta(1+\lambda_{n,s})(I_{f,s}+\sigma^2+(|g_{f,s}|^2+\beta_{k,s}|g_{k,s}|^2|h_{f,k}|^2)p_s)))).$

\hrulefill
\end{figure*}
\begin{alignat}{2}
 & \underset{{\alpha_{n,s},\alpha_{f,s},\beta_{k,s}}}{\text{max}} EE = \sum\limits_{s=1}^S\left(\frac{R_{n,s}+R_{f,s}}{p_s\alpha_{n,s}+p_s\alpha_{f,s}+p_c}\right) \label{10}\\
s.t. &
\begin{cases}
\mathcal C1: \ p_s\alpha_{n,s}\left(|g_{n,s}|^2+\beta_{k,s}|g_{k,s}|^2|h_{n,k}|^2\right)\geq\nonumber\\ \left(2^{R_{min}}-1\right)\left(|g_{n,s}|^2p_s\alpha_{f,s}\delta+I^k_{n,s}+\sigma^2\right),\ \forall s, \\
\mathcal C2: \ p_s\alpha_{f,s}\left(|g_{f,s}|^2+\beta_{k,s}|g_{k,s}|^2|h_{f,k}|^2\right)\geq\nonumber\\ \big(2^{R_{min}}-1\big)\left(p_s\alpha_{n,s}(|g_{f,s}|^2+\beta_{k,s}||g_{k,s}|^2|h_{f,k}|^2\right)\nonumber\\+I_{f,s}+\sigma^2), \forall s, \\
\mathcal C3: \ 0\leq p_s\leq P_{tot},\ \forall s ,\\
\mathcal C4: \ \alpha_{n,s}+\alpha_{f,s}\leq1,\ \forall s,\\
\mathcal C5: \ 0\leq\beta_{k,s}\leq1,\ \forall k,\ \forall s,
\end{cases}
\end{alignat}
where $R_{n,s}=\log_2(1+\gamma_{n,s})$ and $R_{f,s}=\log_2(1+\gamma_{f,s})$ are the data rate of vehicle $n$ and vehicle $f$ associated with RSU $s$. Constraints $\mathcal C1$ and $\mathcal C2$ guarantee the minimum rate of vehicle $n$ and vehicle $f$. Constraint $\mathcal C4$ ensures the power allocation of vehicle $n$ and vehicle $f$ according to NOMA protocol. Constraint $\mathcal C3$ limits the power budget of each RSU while $\mathcal C5$ control the reflection coefficient between 0 and 1.

It can be observed that the optimization problem (4) is coupled on two different variables, i.e., transmit power of RSU and reflection coefficient of backscatter device. This make the problem nonconvex and it is very difficult to get the Global optimal solution. Thus, we provide a two stage solution by decoupling the original problem into two subproblems, i.e., power allocation subproblem and reflection coefficient selection subproblem. First, we optimize the reflection coefficient of backscatter devices in first stage, using the fixed transit power of RSUs. Then, for a given reflection coefficients at backscatter devices, we optimize the transmit power of RSUs at the second stage. 

\subsection{Proposed Energy-efficient Solutions}
Given the transmit power of RSUs, the problem in (3) can be transformed into reflection coefficient optimization as:
\begin{alignat}{2}
 & \underset{{\beta_{k,s}}}{\text{max}} EE = \sum\limits_{s=1}^S\left(\frac{R_{n,s}+R_{f,s}}{p_s\alpha_{n,s}+p_s\alpha_{f,s}+p_c}\right), \label{10}\\
& s.t. \quad
\mathcal C1, \mathcal C2, \mathcal C5.\nonumber
\end{alignat}
Next we use a Lemma to prove that the sum rate in (4) is concave/convex with respect to $\beta_{k,s}$.
\begin{figure*}[!t]
\centering
\includegraphics [width=0.8\textwidth]{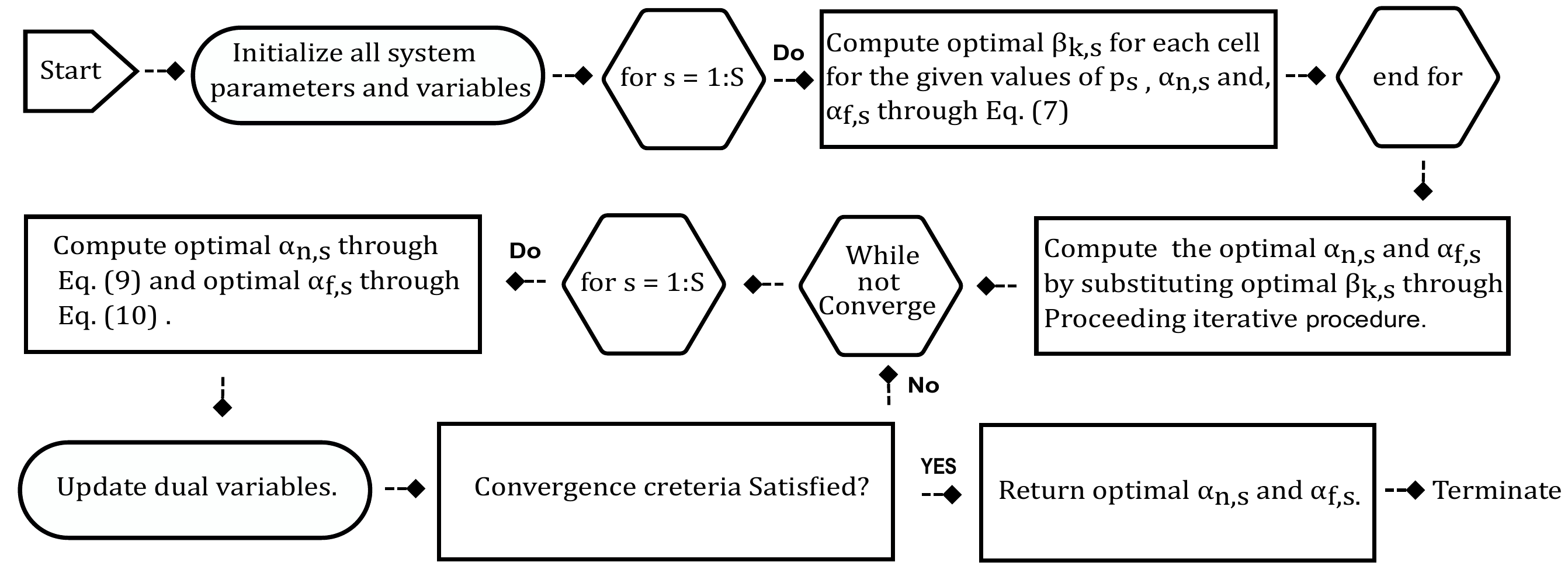}
\caption{Proposed resource optimization Algorithm. After initializing all the variables and parameters, we first calculate the efficient reflection coefficient of each backscatter device for a given power allocation at RSUs. Then, we substitute the calculated values of reflection coefficient and calculate the efficient transmit power at RSUs. This iterative process is terminated when convergence criterion satisfied.}
\label{Alg}\hrulefill
\end{figure*}

\noindent\textbf{\textit{Lemma1:}}
\begin{align}
\hat R_{sum}=R_{n,s}+R_{f,s},\nonumber
\end{align}
is concave/convex function with respect to $\beta_{k,s}$.

\noindent\textbf{\textit{Proof:}} Please refer to Appendix A.

According to Lemma1, $\hat R_{sum}$ is concave/convex, thus, we employ Dinkelbach method which transform the objective function into a subtractive function. By applying this method, the optimization problem in (4) transformed as: 

\begin{alignat}{2}
& \underset{{\beta_{k,s}}}{\text{max}}\ EE  = \underset{{\beta_{k,s}}}{\text{max}}\ F(\vartheta) = \sum\limits_{s=1}^S (R_{sum}-\vartheta P_T)\\
& s.t.\quad\mathcal C1, \mathcal C2, \mathcal C5, \nonumber
\end{alignat} 
where $P_T=p_s(\alpha_{n,s}+\alpha_{f,s})+p_c$ and $\vartheta$ is the real parameter. Moreover, $\vartheta$ represents the maximum energy efficiency such as $\sum\limits_{s=1}^S (R_{sum}-\vartheta P_T)=0$. Note that computing the roots of $F(\vartheta)$ is analogous to solving the objective function in problem (4). It will be negative when $\vartheta$ approaches infinity and positive when $\vartheta$ goes minus infinity. Since the optimization problem in (5) is concave/convex, which motivates us to adopt KKT conditions. After doing straightforward steps, we obtain the closed-form expression as:
\begin{align}
\beta^*_{k,s}= \Bigg[\frac{(2^{R_{min}}-1)p_s\alpha^*_{n,s}|g_{n,s}|^2}{p_s\alpha_{n,s}|g_{k,s}|^2|h_{n,k}|^2}\bigg].
\end{align}
This is the efficient reflection coefficient of backscatter device
k in RSU s (please see Appendix B for the detailed derivations).

By substituting the values of reflection coefficient in problem (3), it is simplified and transformed into a power allocation problem as:
\begin{alignat}{2}
 & \underset{{\alpha_{n,s},\alpha_{f,s}}}{\text{max}} EE = \sum\limits_{s=1}^S\left(\frac{R_{n,s}+R_{f,s}}{p_s\alpha_{n,s}+p_s\alpha_{f,s}+p_c}\right), \label{10}\\
& s.t. \quad
\mathcal C1-\mathcal C4.\nonumber
\end{alignat}
In the following Lemma, we prove that $\hat R_{sum}$ in problem (6) is concave/convex with respect to $\alpha_{n,s}$ and $\alpha_{f,s}$.

\noindent\textbf{\textit{Lemma2:}}
\begin{align}
\hat R_{sum}=R_{n,s}+R_{f,s},\nonumber
\end{align}
is concave/convex function with respect to $\alpha_{n,s}$ and $\alpha_{f,s}$.

\noindent\textbf{\textit{Proof:}} Please refer to Appendix C.

Based on Lemma2, the optimization problem in (6) is concave/convex, thus, Dinkelbach is an efficient tool to get the efficient solution. By using this method, the optimization problem in (6) can be transformed as:
\begin{alignat}{2}
& \underset{{\alpha_{n,s},\alpha_{f,s}}}{\text{max}}\ EE  = \underset{{\alpha_{n,s},\alpha_{f,s}}}{\text{max}}\ F(\vartheta) = \sum\limits_{s=1}^S (R_{sum}-\vartheta P_T)\\
& s.t.\quad\mathcal C1-\mathcal C4.  \nonumber
\end{alignat}
Similar to the solution of problem (5), we exploit dual theory the obtain the efficient solutions. The closed-form solutions for power allocation at RSU $s$ can be derived as:
\begin{align}
\alpha^*_{n,s}=\left[\frac{-y\pm\sqrt{y^2-4xz}}{2x}\right]^+,
\end{align}
\begin{align}
\alpha^*_{f,s}=1-\alpha^*_{n,s},
\end{align}
where $[.]^+=max[0,.]$ and the values of $x$, $y$, $z$ are given at the top of this page. The detailed steps involved in the solution of (8) can be seen in Appendix D.
\begin{figure}[!t]
\centering
\includegraphics [width=0.51\textwidth]{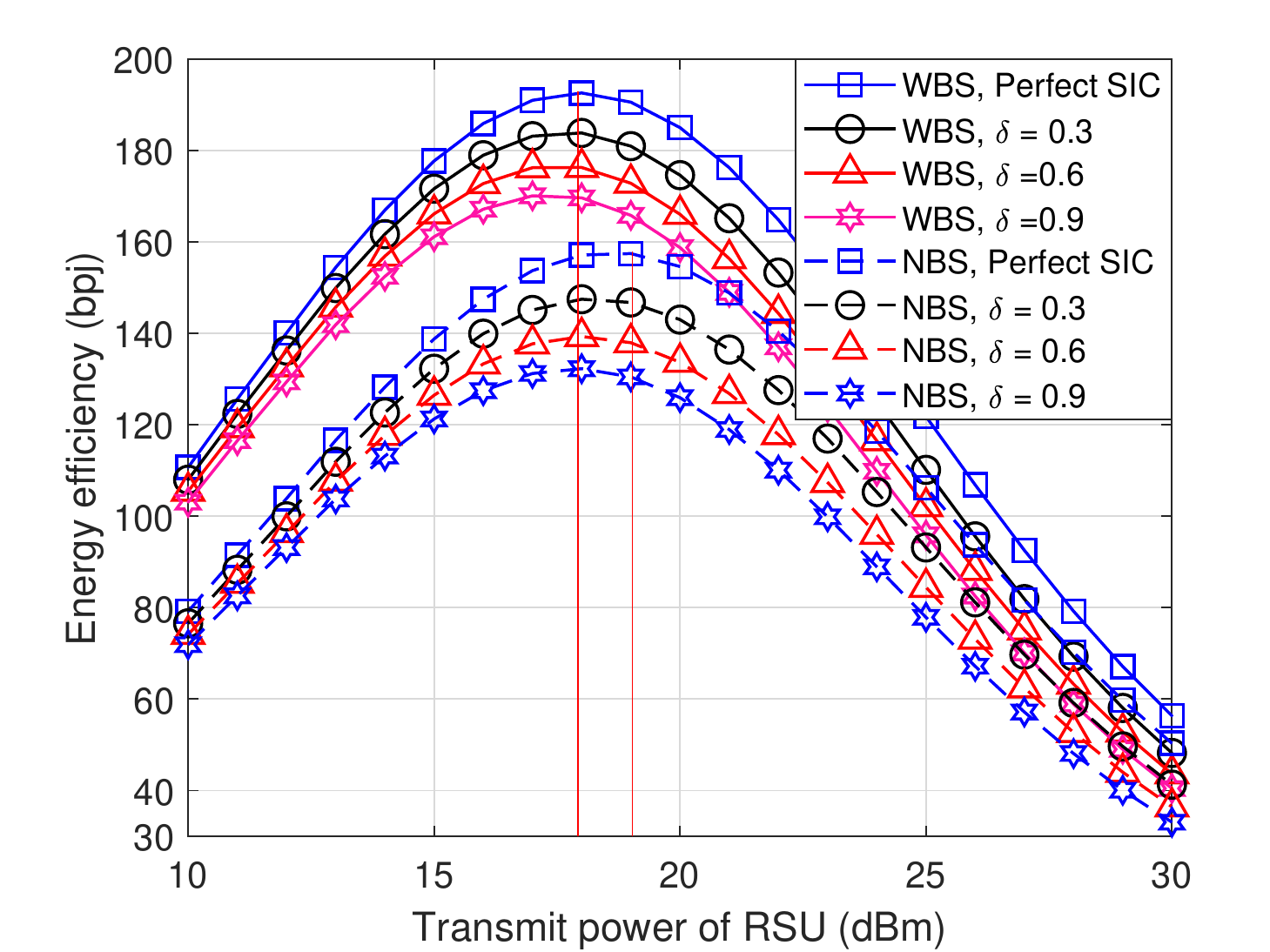}
\caption{Total achievable energy efficiency of the proposed NOMA vehicular network versus the transmit power of each RSU with backscatter communication and without backscatter communication. We plot result for both perfect and imperfect SIC detecting. Here we consider the system with different values of imperfect SIC, i.e., 0.3, 0.6, 0.9. For this figure, we consider the total number of RSUs as $S=10$, the minimum data rate of each vehicle is $R_{min}=0.5$ bps, the transmit power of each RSU varies from $P_{s}=10$ to $P_{s}=30$ dBm, vehicles associated with each RSU are 2, backscatter device in the geographical area of each RSU is 1, and the variance is $\sigma=0.1$.}
\label{Fig5}
\end{figure}

Finally, the Lagrange multipliers involve in the solution process are iteratively updated using sub-gradient method. The proposed iterative algorithm is also shown in Figure \ref{Alg}. It is important to provide the mathematical analysis of the computational complexity for the proposed algorithm. Here the term complexity refer to the number of iteration require for the convergence of the proposed algorithm. It is very very interesting that the complexity of the proposed algorithm increases linearly. For example, if each RSU acommodates two vehicles at one time and the number of total RSUs in the network is $S$. Then the complexity in each iteration can be calculated as $\mathcal O(2)S$. Similarly, if the total iterations are $\Pi$, the total computational complexity will become $\mathcal O\{(2)S\Pi\}$. 

\subsection{Numerical Results and Discussion}
Here we study the numerical results which are obtained using Monte Carlo simulations. We discuss the performance of the considered network with backscattering (WBS) and the benchmark network with no-backscattering (NBS). Unless mentioned, the system parameters are defined as: the total transmit power of each RSU is $P_{tot}=30$ dBm, the total number of RSUs is set as $S=10$, the total number of vehicles is 20 $(n=10$ and $f=10)$, the total number of backscatter devices is 10, the maximum reflection coefficient of each backscatter device is 1, the variance is set as $\sigma^2=0.1$, minimum rate of each vehicle is $R_{min}=0.5$ bps, and the values of imperfect SIC are $\delta=0.1,0.3,0.6,0.9$.
\begin{figure}[!t]
\centering
\includegraphics [width=0.51\textwidth]{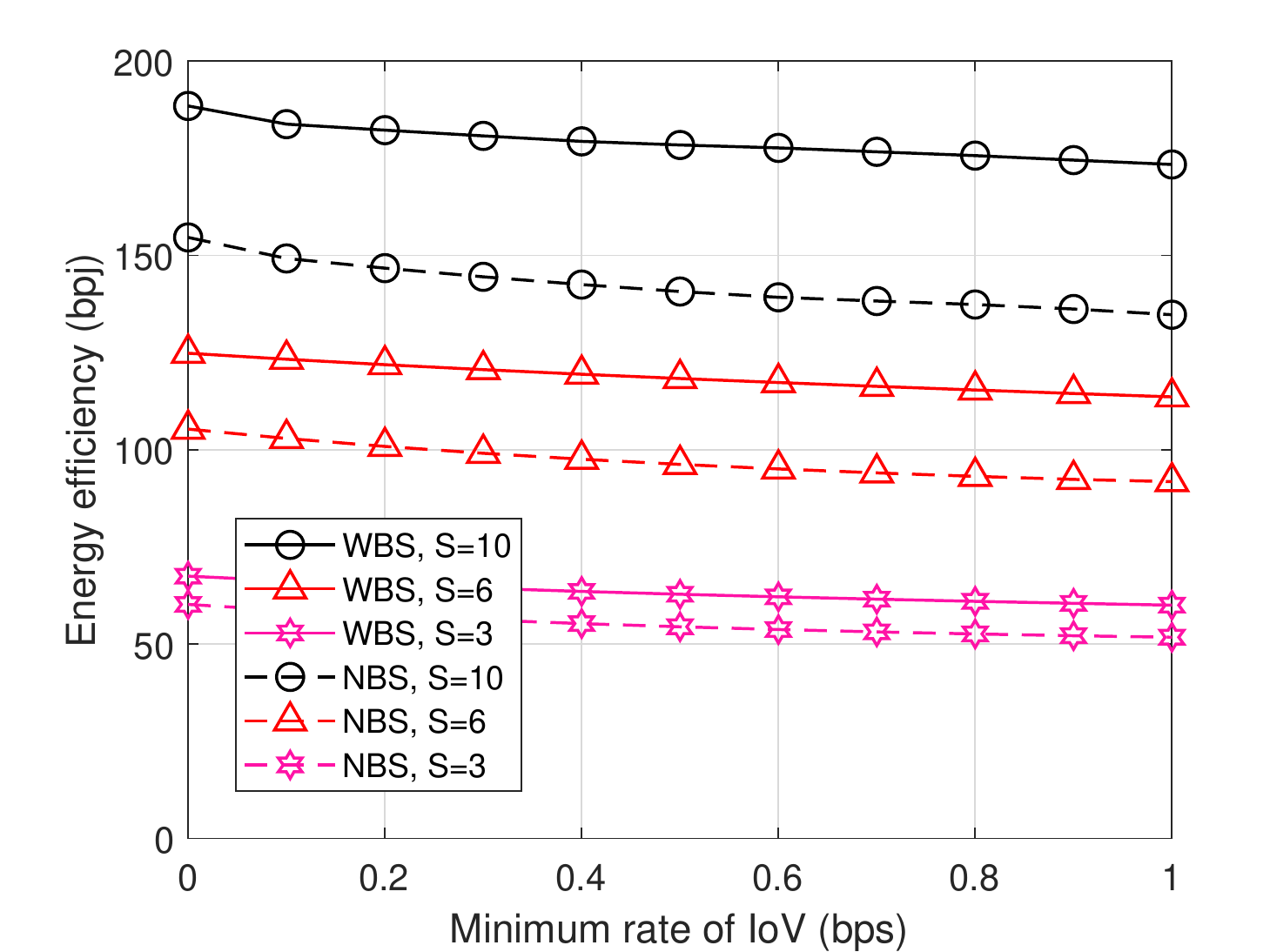}
\caption{Minimum rate of each vehicle versus total achievable energy efficiency of backscatter-enabled NOMA vehicular network with different number of RSUs. Here we consider the system with different RSUs, i.e., 3, 5, 10. For this figure, we consider the value of imperfect SIC parameter as $\delta=0.1$, the minimum data rate of each vehicle varies from $R_{min}=0$ to $R_{min}=1$ bps, the total transmit power of each RSU is $P_{s}=30$ dBm, vehicles associated with each RSU are 2, backscatter device in the geographical area of each RSU is 1, and the variance is $\sigma=0.1$.}
\label{Fig6}
\end{figure}

Figure \ref{Fig5} shows the total achievable energy efficiency against the increasing transmit power of each RSU for both perfect and imperfect SIC detecting cases. For this figure, the values of imperfect SIC parameter are set as $\delta=0.3,0.6,0.9$, and the transmit power of each RSU varies from 10 to 30 dBm. For both WBS amd NBS, the system with perfect SIC detecting provides more energy efficiency as compared to imperfect SIC scenarios. This shows the role of efficient SIC decoding in the successful operation of NOMA systems. Moreover, it is important to note that the total achievable energy efficiency follows
the bell shaped curves in which the efficiency increases with the RSU's transmit power, until a saturating point and then the curves of energy efficiency start dropping with further increase in the transmit power of RSUs. However, we can see that the performance gap between different curves of NOMA-enabled vehicular network for WBS and NBS are significantly large, which shows the advantage of employing backscatter communication in NOMA-enabled IoV networks. It can also be seen that vehicular network with backscatter communication can obtain maximum energy efficiency at lower power budget of RSUs, as compared to the vehicular network scenario without backscatter communication. 

Despite the energy efficiency maximization, it is also important to guarantee the quality of services of NOMA vehicles associated with different RSUs. In this regard, Figure \ref{Fig6} discusses the energy efficiency against the increasing values of minimum rate. In this figure, we set the transmit power as $P_s=30$ dBm, the number of vehicles associated with each RSU is 2, the imperfect SIC parameter is $\delta=0.1$, and the number of RSUs is set as $S=3,6,10$. It can be observed that the achievable energy efficiency decreases as the minimum rate per vehicle increases. However, the proposed NOMA-enabled vehicular network with backscatter communication significantly outperforms the vehicular network withot backscatter communication. We can also see that the achievable energy efficiency increases as the number of cells in the NOMA vehicular network increases. It is because more vehicles can be accommodated by using the same spectrum/time resources. It is also important to mention here that for the small number of RSUs, i.e., $K=3$, the gap of total achievable energy efficiency between WBS and NBS NOMA networks is very small. However, when the number of RSUs is $K=6$ and $K=10$, the gap of achievable energy efficiency between two networks increases significantly. It shows the importance of NOMA with backscatter communication for large-scale vehicular networks.
\begin{figure}[!t]
\centering
\includegraphics [width=0.51\textwidth]{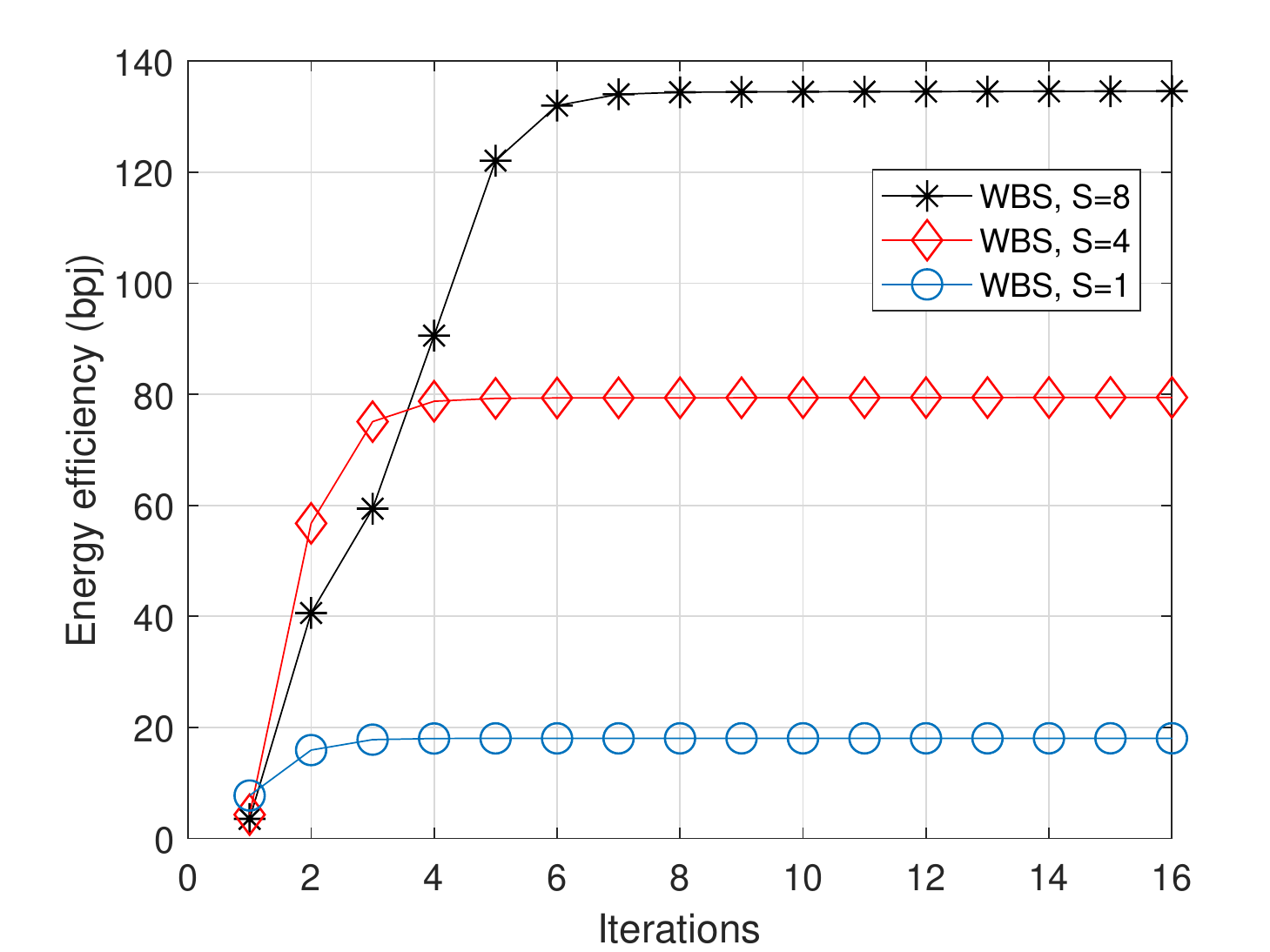}
\caption{Number of iterations requires for the convergence of the proposed algorithm versus total achievable energy efficiency of backscatter-enabled NOMA vehicular network with different number of RSUs. We plot the convergence of the proposed algorithm for the system with different RSUs, i.e., 1, 4, 8. For this figure, we consider the value of imperfect SIC parameter as $\delta=0.6$, the minimum data rate of each vehicle as $R_{min}=1$ bps, the total transmit power of each RSU is $P_{s}=30$ dBm, vehicles associated with each RSU are 2, backscatter device in the geographical area of each RSU is 1, and the variance is $\sigma=0.1$.}
\label{conv}
\end{figure}

Finally, it is important to show the convergence behavior of the proposed algorithm. Figure \ref{conv} shows the achievable energy efficiency ervsus number of iteration with different number of RSUs in the network. Specifically, the energy efficiency is plotted for $S=1,4,8$. It is shown from the figure that the proposed optimization algorithm is low complex and converges after few iterations. We can also observed that the number of iterations required for convergence increases as the number of RSUs in the network increase. For instance, when the number of RSU is 1 and 4, the proposed algorithm converges within 3 and 4 iterations. Comparatively, the same algorithm with high number of RSUs require more iterations for convergence.

\section{Open Issues and Future Research Opportunities}
Here, we cover some of the future opportunities for NOMA and backscatter-enabled 6G vehicular networks.

\subsection{Limited communication Range}
Currently, backscatter communication can be employed in very few V2X scenarios because of its limited communication range. Using V2X for the automotive-industry 5.0, will need reliable data sharing over a long geographical range. Although research is carried out to improve the backscatter communication range through the utilization of power amplifiers, LoRa backscatter, coherent detection for bistatic backscatters, relaying, quantum tunneling, multiple antenna systems, and D2D communications. However, it still needs improvement in its communication range, to meet the QoS demands for the IoV network.

\subsection{Channel estimation}
Sensors in the backscatter-aided IoV system can harvest low power levels through which it modulates its information signal by reflecting the incident signal. Therefore, channel estimation through training signal or pilot signal is not favorable for backscatter communication because it requires high power. Moreover, RF signals are generally unknown to both backscatter device and vehicular users, therefore unpredictability of channel at reflection and absorption stages pose challenges for channel estimation. Therefore, a novel channel estimator is needed.

\subsection{Interference Management}
In 6G V2X communications, various heterogeneous wireless technologies will share the spectrum resources and the network infrastructure. Besides, the spectrum reuse among V2I and V2V links for spectral efficient communication will cause co-channel interference (CCI). Moreover, backscatter aided V2X communications through various passive devices or sensors at the same time will create further interference. Therefore, it demands novel interference cancellation and alignment schemes for interference management.

\subsection{Artificial Intelligence}
Efficient power allocation is a key challenge for integrating backscattering in 6G vehicular networks for industry 5.0. In this regard, efficient machine learning techniques such as reinforcement learning can be developed to improve throughput by reducing interference. Moreover, regression or Artificial Neural Network (ANN) based techniques can be used to predict the RF signal level and energy harvested by the backscatter tag. Also, federated learning can be used to locally compute the power allocation learning models which can be shared with the central server, based on which more accurate global power models can be developed.

\section{Conclusion}
Industry 5.0 will utilize the future 6G communications to enable reliable vehicular networks. 6G will rely on two promising technologies, NOMA and backscatter communications, to provide high data rates and robust connectivity. In this paper, we provide an overview of these two technologies and recent work done in NOMA empowered backscatter communications. We explain three major use cases of backscatter communications in NOMA-enabled 6G vehicular networks. We also present an energy optimization algorithm to highlight the advantages gained by integrating backscatter communications in NOMA-enabled 6G vehicular networks. In the end, we describe the open challenges and future research directions.

Our proposed optimization framework can be easily extended in several ways: 1) This study can be extended two multi-user scenarios. For example, if the cell is partitioned into different clusters and each cluster consists of two vehicles. In this case, vehicles in the same cluster will use NOMA while OMA can be utilized among different clusters. 2) Our work can also be extended to multi-carrier communication in each cell, where each sub-carrier will accommodate two vehicles. 3) We can also investigate the proposed framework under the mobility and imperfect channel state information of the vehicles. These interesting topics will be studied in the future. 

\section*{Appendix A: Proof of Lemma 1}
Here, we proof that problem in (4) is concave/convex for optimization variable $\beta_{k,s}$. To check the concavity/convexity of any function, it is important to calculate its second order derivations. In the following, we first calculate the first order derivations of $\hat R(sum)=R_{n,s}+R_{f,s}$ as: 
\begin{align}
\frac{\partial}{\partial \beta_{k,s}}&\Big[\log_2(1+\gamma_{n,s})+\log_2\big(1+\gamma_{f,s}\big)\Big],\tag{A1}\end{align}
where $\gamma_{n,s}$ and $\gamma_{f,s}$ can be rewritten as:
\begin{align}
\gamma_{n,s}=\frac{\hat X_{n,s}+\hat Y_{n,s}}{\hat Z_{n,s}},\tag{A2}
\end{align}
with $\hat X_{n,s}=p_s\alpha_{n,s}|g_{n,s}|^2$, $\hat Y_{n,s}=p_s\alpha_{n,s}\beta_{k,s}|g_{k,s}|^2|h_{n,k}|^2$, $\hat Z_{n,s}=p_s\alpha_{f,s}|g_{n,s}|^2\delta+I_{n,s}+\sigma^2$, and
\begin{align}
\gamma_{f,s}=\frac{\hat X_{f,s}+\hat Y_{f,s}}{\hat Z_{f,s}+\hat U_{f,s}},\tag{A3}
\end{align}
with $\hat X_{f,s}=p_s\alpha_{f,s}|g_{f,s}|^2$, $\hat Y_{f,s}=p_s\alpha_{f,s}\beta_{k,s}|g_{k,s}|^2|h_{f,k}|^2$, $\hat Z_{f,s}=p_s\alpha_{n,s}\beta_{k,s}|g_{k,s}|^2|h_{f,k}|^2+I_{f,s}+\sigma^2$, and $\hat U_{f,s}=p_s\alpha_{n,s}|g_{f,s}|^2$. With this, (A1) can be stated as:
\begin{align}
\frac{\partial}{\partial \beta_{k,s}}&\Bigg[\log_2\bigg(1+\frac{\hat X_{n,s}+\hat Y_{n,s}}{\hat Z_{n,s}}\bigg)\nonumber\\
&+\log_2\bigg(1+\frac{\hat X_{f,s}+\hat Y_{f,s}}{\hat Z_{f,s}+\hat U_{f,s}}\bigg)\Bigg].\tag{A4}
\end{align}
By computing first order partial derivations, it is expressed as:
\begin{align}
\frac{\partial[\hat R(sum)]}{\partial \beta_{k,s}}&=  \frac {\hat Y_{n,s}}{\ln(2)(\hat X_{n,s}+\beta_{k,s}\hat Y_{n,s}+\hat Z_{n,s})}\nonumber\\&+\frac {\hat Y_{f,s}\hat Z_{f,s}-\hat X_{f,s}\hat U_{f,s}}{\ln(2)((\varOmega)^2+\varOmega(\hat X_{f,s}+\beta_{k,s}\hat Y_{f,s}))},\tag{A5} 
\end{align}
where $\varOmega=\hat Z_{f,s}+\beta_{k,s}\hat U_{f,s}$. Now we compute the second order derivations of (A5) as:
\begin{align}
&\frac{\partial^2}{\partial \beta^2_{k,s}}\bigg[\frac {\hat Y_{n,s}}{\ln(2)(\hat X_{n,s}+\beta_{k,s}\hat Y_{n,s}+\hat Z_{n,s})}\nonumber\\&+\frac {\hat Y_{f,s}\hat Z_{f,s}-\hat X_{f,s}\hat U_{f,s}}{\ln(2)(\hat\varOmega^2_{f,s}+\hat\varOmega_{f,s}(\hat X_{f,s}+\beta_{k,s}\hat Y_{f,s}))}\bigg],\tag{A6} 
\end{align}
After some straightforward steps, it can be written as:
\begin{align}
\frac{\partial^2(A5)}{\partial \beta^2_{k,s}}&=  - \bigg( \frac {{\hat Y^2_{n,s}}}{\ln(2)(\hat X_{n,s}+\beta_{k,s}\hat Y_{n,s}+\hat Z_{n,s})^2} \tag{A7}\\
&+\frac{\hat \varXi_{f,s}(2\hat U_{f,s}\hat A_{f,s}+\hat \varXi_{f,s})}{\ln(2)\hat\varOmega_{f,s}^2({\hat\varOmega_{f,s}}+\hat\varOmega_{f,s}(\hat X_{n,s}+\beta_{k,s}\hat Y_{n,s}))^2}\bigg) < 0\nonumber
\end{align} 
where $\hat A_{f,s}=\hat\varOmega_{f,s}+\hat Y_{f,s}\beta_{k,s}$ and $\hat \varXi_{f,s}=\hat Y_{f,s}\hat Z_{f,s}-\hat X_{f,s}\hat U_{f,s}$. We can see that the second order deviation is less than 0, hence $\hat R(sum)$ is concave/convex and also increasing function.

\section*{Appendix B: Derivations of (6)}
To prove (6), we first define the Lagrangian function of (5) and then calculate its partial derivations. The Lagrangian function of (5) can be expressed as:
\begin{alignat}{2}
& L(\mu_{n,s},\mu_{f,s},\lambda_{s},\tau_{k,s})= \tag{B1}\label{30}\\&\log_2\bigg\{\bigg(1+\frac{\hat X_{n,s}+\hat Y_{n,s}}{Z_{n,s}}\bigg)+\log_2\bigg(1+\frac{\hat X_{f,s}+\hat Y_{f,s}}{\hat Z_{f,s}+\hat U_{f,s}}\bigg)\bigg\}\nonumber\\&-\vartheta p_s(\alpha_{n,s}+\alpha_{f,s})+p_c+\mu_{n,s}(\hat X_{n,s}+\hat Y_{n,s}-\nonumber\\ 
&(2^{R_{min}}-1)\hat Z_{n,s})+\mu_{f,s}({\hat X_{f,s}+\hat Y_{f,s}}-(2^{R_{min}}-1)\times\nonumber\\&({\hat Z_{f,s}+\hat U_{f,s}}))+\lambda_s(P_{tot}-p_s)+ \tau_{k,s}(\beta_{k,s}-1)\nonumber
\end{alignat}
where $\mu_{n,s},\mu_{f,s},\lambda_{s},\tau_{k,s}$, $\mu_m$ and $\eta_{k,m}$ denote the Lagrangian multipliers associated with problem (5). Now we compute the partial deviations of (B1) which can be computed as:
\begin{align}
&\frac{\partial L(\mu_{n,s},\mu_{f,s},\lambda_{s},\tau_{k,s})}{\partial \beta_{k,s}}=\nonumber\\
& \frac {\hat Y_{n,s}}{\ln(2)(\hat X_{n,s}+\beta_{k,s}\hat Y_{n,s}+\hat Z_{n,s})}\nonumber\\ &+\frac {\hat Y_{f,s}\hat Z_{f,s}-\hat X_{f,s}\hat U_{f,s}}{\ln(2)((\varOmega)^2+\varOmega({\hat X_{f,s}+\beta_{k,s}\hat Y_{f,s}})}+\mu_{n,s} \nonumber\\
&\hat Y_{n,s}+\mu_{f,s}(\hat Y_{f,s}-(2^{R_{min}}-1)\hat U_{f,s})+\tau_{k,s} \tag{B2}\label{36}
\end{align}
By letting $\frac{\partial L(\mu_{n,s},\mu_{f,s},\lambda_{s},\tau_{k,s})}{\partial \beta_{k,s}}=0$, it can be written as:
\begin{align}
& \frac {\hat Y_{n,s}}{\ln(2)(\hat X_{n,s}+\beta_{k,s}\hat Y_{n,s}+\hat Z_{n,s})}\nonumber\\ &+\frac {\hat Y_{f,s}\hat Z_{f,s}-\hat X_{f,s}\hat U_{f,s}}{\ln(2)((\varOmega)^2+\varOmega({\hat X_{f,s}+\beta_{k,s}\hat Y_{f,s}})}+\mu_{n,s} \nonumber\\
&\hat Y_{n,s}+\mu_{f,s}(\hat Y_{f,s}-(2^{R_{min}}-1)\hat U_{f,s})+\tau_{k,s} = 0 \tag{B3}
\end{align}

\begin{align}
& \frac {\hat Y_{n,s}}{\ln(2)(\hat X_{n,s}+\beta_{k,s}\hat Y_{n,s}+\hat Z_{n,s})}\nonumber\\ &+\frac {\hat Y_{f,s}\hat Z_{f,s}-\hat X_{f,s}\hat U_{f,s}}{\ln(2)((\varOmega)^2+\varOmega({\hat X_{f,s}+\beta_{k,s}\hat Y_{f,s}})}+\tau_{k,s} \nonumber\\
&=(\mu_{f,s}(2^{R_{min}}-1)\hat U_{f,s}-\mu_{f,s}\hat Y_{f,s})-\mu_{n,s}\hat Y_{n,s}  \tag{B4}
\end{align}
where $\hat Y_{f,s}\hat Z_{f,s}-\hat X_{f,s}\hat U_{f,s} = I_{f,s}+\sigma^2 > 0$ which shows that the left hand side of (B4) is always positive. Hence, the right hand side of (B4) can be given as:
\begin{align}
(\mu_{f,s}(2^{R_{min}}-1)\hat U_{f,s}-\mu_{f,s}\hat Y_{f,s})>\mu_{n,s}\hat Y_{n,s}\tag{B5} \label{38}
\end{align}
here $\mu_{f,s}(2^{R_{min}}-1)\hat U_{f,s}-\mu_{f,s}\hat Y_{f,s}$ is positive because $(2^{R_{min}}-1)$ is positive and $\hat U_{f,s} > \hat Y_{f,s}$. As $\mu_{n,s}$ and $\mu_{f,s}$ are always positive, the slack complimentary criterion of KKT method is always satisfied. Further, the minimum rate constraints associated to $\mu_{n,s}$ and $\mu_{f,s}$ are also active. Thus, both constraints can be equal to 0. Finally, the expression of $\psi_{k,m}$ can be calculated through active inequality constraint as presented in (6).

\section*{Appendix C: Proof of Lemma 2}
Here we prove that problem in (7) is concave/convex for optimization variables $\alpha_{n,s}$ and $\alpha_{f,s}$. To check the concavity/convexity, we will show that the Hessian matrix of the objective function is negative definite. More specifically, the Hessian matrix will be negative definite when its principal minors are alternate in sign. Let us define the hessian matrix first and then calculate its partial derivations as:

\begin{align}
H=
\begin{bmatrix}
\frac {\partial \hat R_{sum}}{\partial^2\alpha_{n,s}} & \frac {\partial \hat R_{sum}}{\partial\alpha_{n,s}\partial\alpha_{f,s}} \\
\frac{\partial \hat R_{sum}}{\partial\alpha_{f,s}\partial\alpha_{n,s}} & \frac {\partial \hat R_{sum}}{\partial^2\alpha_{n,s}}\tag{C1}\label{41}
\end{bmatrix},
\end{align}
Next we calculate the partial deviations of each segment in hessian matrix as:
\begin{align}
\frac {\partial \hat R_{min}}{\partial^2\alpha_{n,s}}\nonumber=Q_{1,1}= -\frac{\hat A_{1,1}-\hat B_{1,1}}{\ln(2)\hat C_{1,1}}, \tag{C2}
\end{align}
where 
\begin{align}
\hat A_{1,1} & =(|g_{n,s}|^2+\beta_{k,s}|g_{k,s}|^2|h_{n,k}|^2)^2((|g_{f,s}|^2\nonumber\\&+\beta_{k,s}|g_{k,s}|^2|h_{f,k}|^2)\alpha_{n,s}+I_{f,s}+\sigma^2)^2.\nonumber\\
\hat B_{1,1} & =(|g_{f,s}|^2+\beta_{k,s}|g_{k,s}|^2|h_{f,k}|^2)^3((|g_{n,s}|^2+\beta_{k,s}|g_{k,s}|^2\nonumber\\ & \times|h_{n,k}|^2)\alpha_{n,s}+\alpha_{f,s}|g_{n,s}|^2\delta)^2(2((|g_{f,s}|^2+\beta_{k,s}|g_{k,s}|^2\nonumber\\&\times|h_{f,k}|^2)\alpha_{n,s}+I_{f,s}+\sigma^2)+(|g_{f,s}|^2+\beta_{k,s}|g_{k,s}|^2\nonumber\\&\times|h_{f,k}|^2)\alpha_{f,s})\alpha_{f,s}.\nonumber\\
\hat C_{1,1} & = ((|g_{n,s}|^2+\beta_{k,s}|g_{k,s}|^2|h_{n,k}|^2)\alpha_{n,s}+\alpha_{f,s}|g_{n,s}|^2\delta)^2\nonumber\\&\times((|g_{f,s}|^2+\beta_{k,s}|g_{k,s}|^2|h_{f,k}|^2)\alpha_{n,s}+I_{f,s}+\sigma^2)^2\nonumber\\&\times((|g_{f,s}|^2+\beta_{k,s}|g_{k,s}|^2|h_{f,k}|^2)\alpha_{n,s}+I_{f,s}+\sigma^2)^2.\nonumber
\end{align}
At the similar way, the partial derivations of second segment can be calculated as:
\begin{align}
&\frac {\partial \hat R_{min}}{\partial\alpha_{n,s}\partial\alpha_{f,s}}=Q_{1,2}=-\frac {\hat A_{1,2}-\hat B_{1,2}}{\ln(2)\hat C_{1,2}}\tag{C3},
\end{align}
where
\begin{align}
\hat A_{1,2} & =(|g_{n,s}|^2+\beta_{k,s}|g_{k,s}|^2|h_{n,k}|^2)|g_{n,s}|^2\delta((|g_{f,s}|^2+\beta_{k,s}\nonumber\\&|g_{k,s}|^2|h_{f,k}|^2)\alpha_{f,s}+(|g_{f,s}|^2+\beta_{k,s}|g_{k,s}|^2|h_{f,k}|^2)\alpha_{n,s})^2.\nonumber\\
\hat B_{1,2} & =(|g_{f,s}|^2+\beta_{k,s}|g_{k,s}|^2|h_{f,k}|^2)^2\nonumber\\ &\times((|g_{n,s}|^2+\beta_{k,s}|g_{k,s}|^2|h_{n,k}|^2)\alpha_{n,s}+I_{n,s}+\sigma^2)^2\nonumber.\\
\hat C_{1,2} & = ((|g_{n,s}|^2+\beta_{k,s}|g_{k,s}|^2|h_{n,k}|^2)\alpha_{n,s}+\alpha_{f,s}|g_{n,s}|^2\delta)^2\nonumber\\&\times((|g_{f,s}|^2+\beta_{k,s}|g_{k,s}|^2|h_{f,k}|^2)\alpha_{n,s}+I_{f,s}+\sigma^2)^2.\nonumber
\end{align}
The partial derivations of third segment can be expressed as:
\begin{align}
&\frac {\partial \hat R_{sum}}{\partial^2\alpha_{f,s}} =Q_{2,1}= - \frac{\hat A_{2,1}-\hat B_{2,1}}{\ln(2)\hat C_{2,1}},\tag{C4} 
\end{align}
where
\begin{align}
\hat A_{2,1} & = (|g_{f,s}|^2+\beta_{k,s}|g_{k,s}|^2|h_{f,k}|^2)^2(|g_{n,s}|^2\delta\alpha_{f,s})^2\nonumber\\&\times((|g_{n,s}|^2+\beta_{k,s}|g_{k,s}|^2|h_{n,k}|^2)\alpha_{n,s}+\alpha_{f,s}|g_{n,s}|^2\delta)^2.\nonumber\\
\hat B_{2,1} & = (|g_{n,s}|^2+\beta_{k,s}|g_{k,s}|^2|h_{n,k}|^2)(|g_{n,s}|^2\delta)^2((|g_{f,s}|^2\nonumber\\&+\beta_{k,s}|g_{k,s}|^2|h_{f,k}|^2)\alpha_{n,s}+I_{f,s}+\sigma^2)^2(2(|g_{n,s}|^2\delta\alpha_{f,s})\nonumber\\&+(|g_{n,s}|^2+\beta_{k,s}|g_{k,s}|^2|h_{n,k}|^2)\alpha_{n,s})\alpha_{n,s} .\nonumber\\
\hat C_{2,1} & = ((|g_{n,s}|^2+\beta_{k,s}|g_{k,s}|^2|h_{n,k}|^2)\alpha_{n,s}+\alpha_{f,s}|g_{n,s}|^2\delta)^2\nonumber\\&\times((|g_{f,s}|^2+\beta_{k,s}|g_{k,s}|^2|h_{f,k}|^2)\alpha_{n,s}+I_{f,s}+\sigma^2)^2\nonumber\\&\times(|g_{n,s}|^2\delta\alpha_{f,s})^2.\nonumber
\end{align}
Finally, the partial derivations of fourth segment can be computed as:
\begin{align}
&\frac {\partial \hat R_{sum}}{\partial\alpha_{f,s}\partial\alpha_{n,s}}=Q_{2,2}=-\frac{\hat A_{2,2}-\hat B_{2,2}}{\ln(2)\hat C_{2,2}},\tag{C5}
\end{align}
where
\begin{align}
\hat A_{2,2}& =(|g_{n,s}|^2+\beta_{k,s}|g_{k,s}|^2|h_{n,k}|^2)(|g_{n,s}|^2\delta)\nonumber\\&((|g_{n,s}|^2+\beta_{k,s}|g_{k,s}|^2|h_{n,k}|^2)\alpha_{n,s}+\alpha_{f,s}|g_{n,s}|^2\delta)^2 .\nonumber\\
\hat B_{2,2} & = (|g_{f,s}|^2+\beta_{k,s}|g_{k,s}|^2|h_{f,k}|^2)^2\nonumber\\&\times(|g_{n,s}|^2+\beta_{k,s}|g_{k,s}|^2|h_{n,k}|^2)\alpha_{n,s}.\nonumber\\
\hat C_{2,2} & = ((|g_{n,s}|^2+\beta_{k,s}|g_{k,s}|^2|h_{n,k}|^2)\alpha_{n,s}+\alpha_{f,s}|g_{n,s}|^2\delta)^2\nonumber\\&\times((|g_{f,s}|^2+\beta_{k,s}|g_{k,s}|^2|h_{f,k}|^2)\alpha_{n,s}+I_{f,s}+\sigma^2)^2.\nonumber
\end{align}
Next the Hessian matrix can be rewritten as:
\begin{align}
H=
\begin{bmatrix}
Q_{1,1} & Q_{1,2} \\
Q_{2,1} & Q_{2,2} \tag{C6}
\end{bmatrix}.
\end{align}
We can observe that the first order principal minors, i.e., $Q_{1,1}$ and $Q_{2,2}$ are negative. And the second order principal minors, i.e., $Q_{1,2}$ and $Q_{2,1}$ are the determinant of the matrix which can be stated as:
\begin{align}
\det H = Q_{1,1}Q_{2,2} - Q_{1,2}Q_{2,1} >0 \tag{C7}
\end{align}
Thus, it proves that the optimization problem in (7) is concave/convex. 

\section*{Appendix D: Derivations of (9)}
Here we calculate the partial derivations of (9) with respect to $\alpha_{n,s}$. The Lagrangian function of (8) can be defined as: 
\begin{alignat}{2}
& L(\mu_{n,s},\mu_{f,s},\lambda_{s},\eta_{s})= \tag{D1}\label{30}\\&\log_2\bigg\{\bigg(1+\frac{\hat X_{n,s}+\hat Y_{n,s}}{Z_{n,s}}\bigg)+\log_2\bigg(1+\frac{\hat X_{f,s}+\hat Y_{f,s}}{\hat Z_{f,s}+\hat U_{f,s}}\bigg)\bigg\}\nonumber\\&-\vartheta p_s(\alpha_{n,s}+\alpha_{f,s})+p_c+\mu_{n,s}(\hat X_{n,s}+\hat Y_{n,s}-\nonumber\\ 
&(2^{R_{min}}-1)\hat Z_{n,s})+\mu_{f,s}({\hat X_{f,s}+\hat Y_{f,s}}-(2^{R_{min}}-1)\times\nonumber\\&({\hat Z_{f,s}+\hat U_{f,s}}))+\lambda_s(P_{tot}-p_s)+ \eta_{s}(1-\alpha_{n,s}-\alpha_{f,s})\nonumber
\end{alignat}
\begin{align}
&\frac{\partial L(\mu_{n,s},\mu_{f,s},\lambda_{s},\eta_{s})}{\partial \alpha_{n,s}}= \nonumber\\&\frac {|g_{n,s}|^2+\beta_{k,s}|g_{k,s}|^2|h_{n,k}|^2}{\ln(2)((|g_{n,s}|^2+\beta_{k,s}|g_{k,s}|^2|h_{n,k}|^2)\alpha_{n,s}+\hat\nu_{n,s})}-\hat\varDelta+\tag{D2}\label{48}\\
&\frac {(|g_{f,s}|^2+\beta_{k,s}|g_{k,s}|^2|h_{f,k}|^2)(|g_{f,s}|^2+\beta_{k,s}|g_{k,s}|^2|h_{f,k}|^2)\alpha_{f,s}}{\ln(2)(\hat\nu_{f,s})((|g_{f,s}|^2+\beta_{k,s}|g_{k,s}|^2|h_{f,k}|^2)\alpha_{f,s}+\hat\nu_{f,s})} \nonumber
\end{align}
where 
\begin{align}
 \hat\nu_{n,s}  = |g_{n,s}|^2\delta\alpha_{f,s}+I_{n,s}+\sigma^2.  \nonumber 
\end{align}
\begin{align}
 \hat\varDelta  =\vartheta p_s-\mu_{n,s}(|g_{n,s}|^2+\beta_{k,s}|g_{k,s}|^2|h_{n,k}|^2)\nonumber\\+\mu_{f,s}(2^{R_{min}}-1)(|g_{f,s}|^2+\beta_{k,s}|g_{k,s}|^2|h_{f,k}|^2)+\eta_s.  \nonumber 
\end{align}
\begin{align}
 \hat\nu_{f,s}  = |(|g_{f,s}|^2+\beta_{k,s}|g_{k,s}|^2|h_{f,k}|^2)\alpha_{f,s}+I_{f,s}+\sigma^2.  \nonumber 
\end{align}
Now we place $\alpha_{f,s}=1-alpha_{n,s}$, (D2) can be stated as:
\begin{align}
&\frac{\partial L(\mu_{n,s},\mu_{f,s},\lambda_{s},\eta_{s})}{\partial \alpha_{n,s}}= \frac {|g_{n,s}|^2+\beta_{k,s}|g_{k,s}|^2|h_{n,k}|^2}{\ln(2)(\hat \varLambda_{n,s}\alpha_{n,s}+\hat \varPsi_{n,s})}\nonumber\\& -\frac {\gamma_{f,s}(|g_{f,s}|^2+\beta_{k,s}|g_{k,s}|^2|h_{f,k}|^2)}{\ln(2)(\hat \varOmega_{f,s}\alpha_{n,s}+\hat \varPsi_{f,s})} -\hat \varDelta\tag{D3}
\end{align}
where 
\begin{align}
\hat \varLambda_{n,s}=(|g_{n,s}|^2+\beta_{k,s}|g_{k,s}|^2|h_{n,k}|^2)-(|g_{n,s}|^2\delta).\nonumber
\end{align}
\begin{align}
\hat \varOmega_{f,s}&=(|g_{f,s}|^2+\beta_{k,s}|g_{k,s}|^2|h_{f,k}|^2)\nonumber\\&-(|g_{f,s}|^2+\beta_{k,s}|g_{k,s}|^2|h_{f,k}|^2).\nonumber
\end{align}
\begin{align}
\varPsi_{n,s}= |g_{n,s}|^2\delta+I_{n,s}+\sigma^2.\nonumber
\end{align}
\begin{align}
\varPsi_{f,s}=(|g_{f,s}|^2+\beta_{k,s}|g_{k,s}|^2|h_{f,k}|^2)+I_{f,s}+\sigma^2.\nonumber
\end{align}
After $\frac{\partial L(\mu_{n,s},\mu_{f,s},\lambda_{s},\eta_{s})}{\partial \alpha_{n,s}}=0$ and some straightforward calculation, it can be expressed as:
\begin{align}
&(|g_{n,s}|^2+\beta_{k,s}|g_{k,s}|^2|h_{n,k}|^2)(\hat \varOmega_{f,s}\alpha_{n,s}+\hat\varPsi_{f,s})\nonumber\\&\gamma_{f,s}(|g_{f,s}|^2+\beta_{k,s}|g_{k,s}|^2|h_{f,k}|^2)(\hat\varLambda_{n,s}\alpha_{n,s}+\hat\varPsi_{n,s})\nonumber\\&-\ln(2)\hat\varDelta_{n,s}(\hat \varOmega_{f,s}\alpha_{n,s}+\hat\varPsi_{n,s})(\hat \varLambda_{n,s}\alpha_{n,s}+\hat\varPsi_{n,s})=0.\tag{D4}
\end{align}
Next we use quadratic formula to rearrange (D4) as:
\begin{align}
&(-\ln(2)\hat \varDelta_{n,s}\hat \varLambda_{n,s}\hat \varOmega_{f,s})\alpha_{n,s}^2+((|g_{n,s}|^2+\beta_{k,s}|g_{k,s}|^2|h_{n,k}|^2)\nonumber\\ &\hat\varOmega_{f,s}-{\gamma_{f,s}(|g_{f,s}|^2+\beta_{k,s}|g_{k,s}|^2|h_{f,k}|^2)}\hat \varLambda_{n,s}\nonumber\\
&-\ln(2)\hat\varDelta_{n,s} \hat \varLambda_{n,s}\hat \varPsi_{f,s}-\ln(2)\hat \varDelta_{n,s}\hat \varOmega_{f,s}\hat \varPsi_{n,s})\alpha_{n,s}+\nonumber\\
&((|g_{n,s}|^2+\beta_{k,s}|g_{k,s}|^2|h_{n,k}|^2)\hat \varPsi_{f,s}-(\gamma_{f,s}(|g_{f,s}|^2\nonumber\\ &+\beta_{k,s}|g_{k,s}|^2|h_{f,k}|^2)\hat \varPsi_{n,s}-\ln(2)\hat \varDelta \hat \varPsi_{n,s}\hat \varPsi_{f,s})\tag{D5} 
\end{align}
Finally, $\alpha_{n,s}$ can be computed as:
\begin{align}
\alpha^*_{n,s}=\left[\frac{-y\pm\sqrt{y^2-4xz}}{2x}\right]^+,
\end{align}
Thus, the derivations of (9) is proved.

\ifCLASSOPTIONcaptionsoff
\newpage
\fi

\bibliographystyle{IEEEtran}
\bibliography{Wali_EE}

\end{document}